\documentstyle[12pt, epsf]{article}
\newcommand{\beq}{\begin{equation}}   
\newcommand{\eeq}{\end{equation}}

\newcommand{\gsim}{\lower.7ex\hbox{$
\;\stackrel{\textstyle>}{\sim}\;$}}
\newcommand{\lsim}{\lower.7ex\hbox{$
\;\stackrel{\textstyle<}{\sim}\;$}}

\begin{document}
\begin{titlepage}
\renewcommand{\thefootnote}{\fnsymbol{footnote}}

\begin{center} \Large
{\bf Theoretical Physics Institute}\\
{\bf University of Minnesota}
\end{center}
\begin{flushright}
TPI-MINN-96/07-T\\
UMN-TH-1428-96\\
hep-ph/9606281
\\
\end{flushright}
\vspace{.3cm}
\begin{center}
{ \Large
{Little Miracles of  Supersymmetric Evolution of  Gauge 
Couplings}} 

\vspace{1cm}

{Talk \footnote{Extended version} at the 4th International Workshop 
on 

{\it Supersymmetry and Unification of Fundamental Interactions}

SUSY-96 (May 29 -- June 1)

University of Maryland, College Park, MD 20742, USA}
   
\end{center}
\vspace*{.3cm}
\begin{center} {\Large 
M. Shifman} \\
\vspace{0.4cm}
{\it  Theoretical Physics Institute, Univ. of Minnesota,
Minneapolis, MN 55455}
\end{center}

\vspace*{.2cm}
\begin{abstract}

The invention of supersymmetry, almost exactly 25 years ago
\cite{GL}, changed the face of  high-energy physics. The idea that the 
observed low-energy gauge groups appear due to the process of 
spontaneous breaking of a single unifying group $G$ is also quite 
popular. The synthesis of these two elements results in 
supersymmetric grand unification. I present (perturbatively) exact 
results regarding the supersymmetric evolution of the gauge 
couplings from the scale of their unification to lower scales.
In particular, it is shown how the heavy mass thresholds can be 
properly taken into account to all orders. 
 
\end{abstract}
\end{titlepage}

When I was kindly invited by Prof. Mohapatra, almost a year ago, to
give this talk, it was supposed that the main subject would be 
the $\alpha_s$ crisis \cite{Shif}. Since then the situation with
low versus high-energy determinations of $\alpha_s(M_Z)$
was reviewed several times \cite{Shif1}, and nothing exciting 
happened during this year. The irreconcilable contradiction between 
the values
of  $\alpha_s(M_Z)$ from the low-energy measurements and the 
high energy determinations is not going to disappear -- this fact 
gradually becomes obvious to everybody.  Perhaps, it is worth 
mentioning a very recent record-breaking  analysis of deep inelastic 
scattering \cite{KKPS} which achieves three-loop accuracy, for the 
first time ever in this
problem. The analysis  confirms, with better  stability,
the earlier conclusion, that $\alpha_s(M_Z)$ is equal to 0.110 with 
the uncertainty close to 0.005. According to Ref. \cite{KKPS}, this is a 
conservative estimate of the error bars.  If we compare this result 
with the values of $\alpha_s$ extracted from the measurements at 
the $Z$ peak we observe a discrepancy of 2.5 to 3 standard 
deviations. Thus, the question is not whether or not new physics is 
already with us but, rather, what kind of new physics manifests 
itself in the existing data. Quite a few talks at this 
Workshop were devoted  to 
various new physics scenarios, expressing various degrees of 
confidence in the scenarios under consideration \cite{Kane}. 
I have nothing to add in this respect. In the absence of fresh ideas 
concerning the $\alpha_s$ crisis, I decided to discuss today another 
topic which, on one hand,  is rather closely related to the
$\alpha_s$ problem and, on the other hand, did not receive
due attention previously -- a very peculiar pattern of the 
supersymmetric evolution of the gauge couplings. 

\section{Outlining the Problem}

The bold idea of grand unification, after two decades of its existence 
\cite{GG}, has become an indispensable part of phenomenology. It 
has 
obviously attractive features, e.g.   the fact that it naturally explains 
the electric charge quantization. Its most unpleasant  feature, to my 
mind,  is the 
prediction of a great desert above the supersymmetry scale
(100 GeV range) up to the grand unification scale ($10^{16}$ GeV 
range), an energy interval spanning fourteen orders of magnitude  or 
so, where nothing happens. If the great desert does indeed exists, 
high energy physics will practically cease to exist as the 
experimental 
science we know today, after the discovery of the superpartners.  I 
would not like  to believe in such a  frightening future, and, 
yet, I am going to discuss the supersymmetric grand unification, for 
two reasons. First, I will 
 emphasize   some general aspects of  supersymmetric 
evolution of the gauge couplings which may well be of importance in 
other problems, not necessarily related to geography of  the great 
desert. 
Treating the gauge coupling evolution in a standard way 
\cite{SSGU},
one overlooks elegant properties which are specific to 
supersymmetry. My task is to reveal these elegant properties. 
Second, supersymmetric grand unification {\it per se}
is an intellectual game occupying the minds of many, probably, 
because it predicts a relation between the low-energy gauge 
couplings which is surprisingly close to the one observed 
experimentally.  It seems worth joining the club of unifiers merely to 
understand 
whether the experimentally observed relation is actually reproduced 
in the 
most naive and straightforward version of supersymmetric grand 
unification. 

In the first part of my talk I will focus on the generalities of the 
supersymmetric evolution of the gauge couplings. 
Assume that there is a (simple)  gauge group $G$ which is 
spontaneously broken down to several subgroups. Far above the 
symmetry breaking scale we start from  a single gauge coupling.
Then it splits, and far below the breaking scale we deal with a whole 
set of independent couplings which, however, remember their 
common origin.  The common origin imposes certain constraints. 

There is nothing new in this formulation, of course. Getting 
observable 
consequences requires evolving the constants from the unification 
scale down to a lower scale -- this fact was recognized almost 
immediately \cite{GQW} after the emergence of the grand
unified theories (GUT).  Huge amounts of literature are devoted to 
this issue.
Let me mention early works on supersymmetric grand unification
\cite{earlySSGU}, and more recent famous works \cite{recGU}
which essentially buried (at least, psychologically)
non-supersymmetric unification. All analyses of the supersymmetric 
evolution of which I am aware  borrow the old technique used in the
non-supersymmetric case \cite{GQW}.  A new element which I would 
like to promote is the full exploitation of  very specific features of
the supersymmetric $\beta$ functions \cite{NSVZ}. Among other 
virtues, this new approach allows one to exactly solve ({\it to all 
orders})
the problem of the  supersymmetric heavy thresholds. 

Supersymmetric 
grand unification, in the context of the realistic model-building, is 
briefly
discussed in the second part. Here some simple numerical estimates 
are presented explaining the impact of various effects on    $\alpha_s 
(M_Z)$. 

\section{Full Beta Function from One Loop of Perturbation Theory}

First, I would like to explain how the full multiloop (perturbative) 
$\beta$ function
can be obtained from a trivial {\em one-loop} calculation. The 
derivation  presented below was reported  over a decade ago 
\cite{VZS}. Since this paper is hardly 
known to anybody it
is instructive to reiterate the main points, the more so 
that we will need  them later on, in discussing the threshold 
effects.

Let us consider a generic non-Abelian gauge theory with the matter 
sector which includes (i) one chiral superfield in the adjoint 
representation of the gauge group (ii) additional matter superfields 
in arbitrary representations. For simplicity I will  keep in mind that 
the gauge group $G$ is $SU(N)$ and assume that the additional 
matter 
is such that a mass term is possible for every superfield. Both 
assumptions can be lifted; moreover, the final expression we will 
obtain shortly is absolutely general. 

In more or less standard (and somewhat symbolic) notation the 
action has the form
$$
{\cal S}_0 =
\frac{1}{2g_0^2} {\rm Tr}\, \int d^2\theta d^4 x W^2 
+\frac{1}{2g_0^2}{\rm Tr}\, \int d^2\theta d^2\bar \theta d^4 x 
\bar\Phi e^{V} \Phi 
$$
\beq
+ \frac{1}{4} \int d^2\theta d^2\bar \theta \sum_f
{S_f}^\dagger e^{V}S_f
+
\left( \frac{1}{4} m_f^0 \int d^2\theta (S_f)^2 +{\rm h.\,  c.}\right)
\label{action}
\eeq
where $g_0$ is the bare coupling constant, $\Phi$ is the superfield in 
the adjoint representation,
$$
\Phi \equiv \sum_{a=1}^{N^2-1}\Phi^a T^a,
$$
 $S_f$ is the set of all other superfields, $f$ is the ``flavor" index, and
$m_f^0$ is the bare mass term. Finally, $T^a$ are $N\times N$ 
traceless matrices, generators of $SU(N)$. Let us note that the 
normalizations of the matter fields are different -- $\Phi$ includes 
the coupling constant while $S_f$'s do not. 
As will become clear shortly, this is convenient.
The bare $Z$ factors of all 
fields $S_f$ are set equal to unity. The lowest component of the 
superfield $\Phi$ will be denoted by $\varphi$. 

One may add all conceivable (super) Yukawa interactions between 
$S_f$; this does not change our argument. Perhaps, it is worth 
noting that in the absence of the 
additional $S_f$ matter fields we actually deal with the extended 
$N=2$ supersymmetry.

The model described by the action (\ref{action}) possesses $D$-flat 
directions -- a system of degenerate classical vacua in the space of 
$\Phi$ fields , the so called valleys;
the potential energy along the bottom of the valley vanishes.
The valleys are parametrized by $N-1$ chiral invariants,
Tr $\Phi^2$,  Tr $\Phi^3$, ... , Tr $\Phi^{N-1}$. Alternatively, one can 
say that   for 
any diagonal matrix  $\Phi$ the $D$ terms vanish. For instance, for 
$SU(3)$ choosing $\varphi^3$ = a complex constant and $\varphi^8$ = 
a 
complex constant and putting all other $\varphi$'s to zero we get a  
point belonging to  the bottom of the valley. 

There is only one point, namely the origin ($\varphi^a = 0$ for all 
$a$),
where the full original gauge symmetry is unbroken. Any other point
from the bottom of the valley corresponds to the spontaneous 
breaking of the symmetry. (This statement is valid, generally 
speaking, only in perturbation theory. Remember,  however, that
I am discussing only perturbation theory). Generically, $SU(N)$ is 
broken 
down to $U(1)^{N-1}$. As a result, we have $N-1$ 
massless ``photons" with their superpartners, and $N^2 - N$ massive
``$W$ bosons", with their superpartners. These ``$W$ bosons" are 
analogs of the superheavy $X$ and $Y$ bosons of the standard theory 
of grand 
unification, sometimes called 
elephants. (Quite naturally -- elephants live in the areas at 
the far side of the desert! All gauge bosons of the group $G$ which 
acquire masses through the expectation value of the scalar field 
belonging to  the adjoint representation of $G$ will be referred to 
below as elephants.)
 Their masses 
need not be equal to each other. For example, for $SU(3)$
$$
M_{1,2}^2 =|\varphi^3|^2,
$$
\beq
M_{4,5}^2 =\frac{1}{4}|\sqrt{3}\varphi^8  + \varphi^3|^2,\,\,\,\,
M_{6,7}^2 =\frac{1}{4}|\sqrt{3}\varphi^8  - \varphi^3|^2 \, .
\label{masses}
\eeq

Now, we  start our  evolution from the normalization point $\mu$ 
equal to the 
ultraviolet cut 
off $M_0$, where the action coincides with the bare one, Eq. 
(\ref{action}), and then descend down  from 
$M_0$   
to some low normalization point.  $M_0$ is assumed to be much 
larger than 
any other scale in the problem at hand. As we cross the mass 
thresholds we
integrate out the corresponding massive fields, one after another.
In this way we integrate out the elephants and all additional matter 
fields $S_f$. 
In the low-energy limit the only surviving  fields are the $N-1$ 
massless 
photons (and their superpartners), which are mutually neutral with 
respect to each other and do not interact. Since the residual gauge 
symmetry 
is $U(1)^{N-1}$ and there is no charged matter left  below the 
last threshold, the gauge couplings do not run below the 
last threshold.  There are $N-1$ 
low-energy gauge couplings corresponding to $N-1$  unbroken 
$U(1)$ subgroups. These  charges obtained after the completion of 
the evolution will be referred to as 
low-energy (frozen)  charges $\alpha$. They could be measured in a 
{\em gedanken} experiment, say,  through the Coulomb interaction
of heavy probe charged particles at large distances.

We are free to choose any low-energy (frozen)  charge and consider 
it as a 
function of the 
ultraviolet cut off, the bare coupling constant $\alpha_0 \equiv 
g_0^2/4\pi$  and other parameters of the 
theory. Varying $M_0$ and $\alpha_0$ in a concerted way, ensuring 
that the low-energy quantities stay fixed, we find the $M_0$ 
dependence of $\alpha_0$ which is equivalent to the knowledge of 
the $\beta$ function. For definiteness, one may  concentrate on  the 
charge corresponding to 
the ``third photon", $A_\mu^3$, although this specific choice is of no 
importance for our purpose in this part. 

To connect $\alpha$ and $\alpha_0$ one needs an explicit formula 
for the low-energy effective action,
\beq
{\cal S} = \frac{1}{4}\left(\frac{1}{g^2}\right)_{ij}\int d^2\theta d^4 x 
\,\,
W^iW^j
 +\frac{Z_{ij}}{4g_0^2} \int d^2\theta d^2\bar \theta d^4 x \,\,
\bar\Phi^i 
\Phi^j \, .
\label{effaction}
\eeq 
This action includes only photons and the neutral adjoint matter. 
Correspondingly, 
the indices $i,j$ run over the Cartan subalgebra (3 and 8 in 
the $SU(3)$
example; in general they take rank$(G)$ different values), and 
$Z_{ij}$ is the 
set of renormalization $Z$ factors of the neutral adjoint matter fields
evolved down ``to the very end".  Note that the kinetic term of the
low-energy photons need not be diagonal in $i$ and $j$. 

Limiting ourselves to the third photon gauge coupling ($i=j=3$) one 
can 
trivially
write the expression for $1/g^2$ at one loop,
\beq
\frac{1}{g^2} 
=\frac{1}{g_0^2}
-\frac{2N}{8\pi^2} \ln\frac{M_0}{\tilde\varphi_0} +\sum_f 
\frac{T(R_f)}{8\pi^2}\ln\frac{M_0}{m_f^0}\, . 
\label{alpha}
\eeq
Here
$T(R_f)$ is (one half of) the Dynkin index,
$$
{\rm Tr}\, (T^a T^b) = T(R)\delta^{ab},
$$
$T^a$ are the group generators in the representation corresponding 
to the 
superfield $S_f$. For the adjoint representation of $SU(N)$
the corresponding index $T({\rm adjoint})=T(G)=N$. This fact is 
actually used in Eq.
(\ref{alpha}); $2N$ in front of the first logarithm  is 
actually $2T(G)$. For the fundamental representation of
$SU(N)$ the index $T({\rm fund}) = 1/2$ for each chiral superfield.
It is worth reminding that one flavor requires two chiral superfields 
in this case, so that effectively $T=1$ for each flavor. Finally, 
$\tilde\varphi$ is a homogeneous function of the moduli whose 
concrete form depends on the group under consideration. In the 
$SU(3)$ case for the third photon
$$
\tilde\varphi = \left( \varphi^3\right)^{2/3}
\left[\frac{1}{4}\left( \varphi^3\right)^{2}-
\frac{3}{4}\left( \varphi^8\right)^{2}\right]^{1/6}\, .
$$
This rather clumsy expression is nothing but   a combination of 
masses (\ref{masses}).  It is important to note that
$\tilde\varphi$ depends only on $\varphi$, not on
$\varphi^\dagger$, and the only singular points (i.e. those where
$\tilde\varphi=0$) are
\beq
\varphi^3 = 0\,\,\, {\rm or}\,\,\,  \varphi^3 =\pm \sqrt{3}\varphi^8\, 
.
\label{sing}
\eeq
At these points the pattern of the symmetry breaking is not generic;
instead of $SU(3)\rightarrow U(1)^2$ we have $SU(3)\rightarrow 
U(1)\times SU(2)$, there are massless non-neutral gauge bosons,
and the notion of the low-energy frozen action becomes ill-defined. 
We will 
stay away from the singular points.

Now we come to  a miracle. Although the low-energy coupling 
constant
(\ref{alpha}) was obtained at one loop this expression is 
(perturbatively) exact to all orders -- the gauge part of the effective 
action 
(\ref{effaction}) 
receives no correction at the level of two loops, three loops, and so on 
-- at every finite order. This non-renormalization theorem stems 
from the fact that {\it if the theory is fully regularized in the 
infrared 
domain} there are no holomorphic anomalies \cite{SV1}, and the 
coefficient in front of $W^2$ in the effective action must be an 
analytic function
of $g_0^{-2}$, $m_f^0$ and the moduli fields \cite{CFGP,SV2,Seiberg}.
In the context of the string theory a similar observation was made in 
Ref. \cite{DKL}.
A more technical proof relying on specific properties of
the background field technique for supergraphs was given in 
\cite{SV2} (see also \cite{SV3}) where the reader will also find a 
comprehensive list of
earlier works in this direction.
 Note 
that the vacuum expectation values of the fields $\varphi$ in all 
expressions above are
those of the bare fields -- the assertion of the holomorphic 
dependence refers to the {\em bare} parameters. Had  the coefficient 
in 
front of $W^2$  been expressed in terms of the renormalized 
(low-energy) parameters, the holomorphicity would be lost. To be 
consistent, I should have marked the field $\tilde\varphi$ by the 
subscript 
``0"; I will do this occasionally, but not in every expression.  The 
proliferation of indices makes them unreadable, so sometimes we 
will just keep  this subscript in mind.

It is very easy to see why higher corrections can not appear. 
Say, the two-loop term in the coefficient of $W^2$, if it existed, 
should be proportional to $1/({\rm Re}\,  g_0^{-2})$, since the 
imaginary part of $g_0^{-2}$ (the $\theta$ term) can not show up 
in the perturbative expansions. In which case 
there is no way to maintain the holomorphic dependence.

Let us pause here and reiterate.
The relation between the physical gauge coupling and the bare one 
presented above is
(perturbatively) exact. The left-hand side is required to be 
independent of the ultraviolet cut off. This requirement determines 
the $M_0$ dependence of $g_0^{-2}$ and, hence, the full  $\beta$ 
function for the gauge coupling. 

The $\beta$ 
function is defined as
$$
\beta (\alpha_0) = {d\alpha_0}/{d\ln M_0}\, .
$$
Naively, one might conclude that the one-loop expression
for the physical charge (\ref{alpha}) yields the one-loop $\beta$ 
function. Everybody knows that this  conclusion is  wrong, of 
course.  So, there should be a subtlety. The question is ``where ?". 

Within our formulation the change  of $g_0^{-2}$ under the variation 
of $M_0$
should be adjusted in such a way as to keep {\it all} physical 
low-energy
 parameters fixed, not only $g^{-2}$.  In particular,
the physical (renormalized) masses of the matter fields  and of the 
elephants
must stay fixed; $\tilde\varphi$ and $m_f$ differ from
$\tilde\varphi_0$ and $m_f^0$, however,  by the corresponding $Z$ 
factors,
which, in turn, do depend on $M_0$. Therefore, differentiating
the right-hand side of Eq. (\ref{alpha}) over $\ln M_0$ 
it is necessary to differentiate $\tilde\varphi_0$ and $m_f^0$
as well. 

It is convenient to rewrite Eq. (\ref{alpha}) 
as follows
$$
\frac{1}{g^2} 
=\left( \frac{1}{g_0^2}
-\frac{2N}{8\pi^2} \ln\frac{M_0}{\tilde\varphi} +\sum_f 
\frac{T(R_f)}{8\pi^2}\ln\frac{M_0}{m_f}\right)
$$
\beq
+\left( 
\frac{2N}{8\pi^2}\ln\frac{\tilde\varphi_0}{\tilde\varphi}+\sum_f
\frac{T(R_f)}{8\pi^2}\ln\frac{m_f}{m_f^0}\right)\, .
\label{alphaprim}
\eeq
The sum of the first and the second brackets is holomorphic, each of 
them separately is not. The non-holomorphicity creeps in at this 
stage 
because the $Z$ factors connecting, say, $m_f$ and $m_f^0$ are not 
holomorphic. The second line in Eq. (\ref{alphaprim}) reduces
to 
\beq
-\frac{1}{4\pi}\left( 
\frac{N}{2\pi}\ln\frac{Z\alpha}{\alpha_0}+\sum_f
\frac{T(R_f)}{2\pi}\ln Z_f\right)\, .
\label{Z}
\eeq
In the absence of the additional matter fields supersymmetry 
extends to $N=2$, and the $Z$ factor of the adjoint matter
is $Z=\alpha_0/\alpha$. 
In other words, $\tilde\varphi$ is not renormalized in this case, and 
the
$\beta$ function is obviously one-loop. 

In deriving Eq. (\ref{Z})
I used the fact that the mass term of the additional matter, being the 
$F$ term, is not renormalized
\cite{NRT}, and the entire renormalization of the mass of the $f$-th 
flavor comes from the corresponding $Z_f$ factor. With the additional 
matter 
fields
included, the $N=2$ supersymmetry degrades  to $N=1$, and 
Eq. 
(\ref{Z})
generates the second and all higher-order terms in the $\beta$ 
function.
Indeed, differentiating $\alpha_0$ over $\ln M_0$ it is trivial to  get
$$
0 = -\frac{1}{\alpha_0^2}\beta ({\alpha_0}) - \frac{2N}{2\pi}
+\sum_f\frac{T(R_f)}{2\pi}\,  +
$$
\beq
\frac{N}{2\pi}\frac{1}{\alpha_0}\, \beta{\alpha_0}
-\left( \frac{N}{2\pi}\gamma+\sum_f
\frac{T(R_f)}{2\pi}\gamma_f\right)
\label{prebeta}
\eeq
where $\gamma$ is the anomalous dimension of the adjoint matter,
$\gamma_f$ is that of $S_f$,
$$
\gamma =d\,\ln Z /d\, \ln M_0
$$
and for simplicity it is assumed that there is no flavor mixing in 
$Z_f$'s
due to (possible) Yukawa interactions, which may be present, in 
principle, although I have not indicated them explicitly in Eq. 
(\ref{action}). 
 Needless to say that this 
assumption is not crucial.
The first and second lines in Eq. (\ref{prebeta}) 
come 
from the
first and second lines in Eq. (\ref{alphaprim}), respectively. 
Combining various 
pieces together we arrive at
\beq
\beta (\alpha ) =
-\frac{\alpha^2 }{2\pi}\left(1-\frac{N\alpha}{2\pi}\right)^{-1}
\left[ 2N -\sum_f T(R_f) +N\gamma + \sum_f 
T(R_f)\gamma_f\right]\, .
\label{beta}
\eeq
This  is nothing but a particular case (occurring when one of the 
matter 
fields is in the adjoint representation) of the general NSVZ $\beta$ 
function
\cite{NSVZ}
\beq
\beta (\alpha ) =
-\frac{\alpha^2 }{2\pi}\left(1-\frac{T(G) \alpha}{2\pi}\right)^{-1}
\left[ 3T(G) -\sum_i T(R_i)(1-\gamma_i)\right]
\label{NSVZbeta}
\eeq
where in the expression above the sum on the right-hand side runs 
over {\em 
all} matter fields \cite{exact}.

\section{Where have the higher orders gone? }

Now,  that  I've explained  how the multiloop $\beta$ function 
appears, I am going to tell you that essentially it is never needed
{\it per se}. 
Our task is calculating the gauge coupling; this calculation requires 
integration of the $\beta$ function. But if we know the full 
expression for the gauge couplings beforehand why bother about 
getting the $\beta$ function and then integrating it back to find the 
couplings?

If the evolution of the gauge couplings is complete and they are 
frozen,
at the very ``bottom", one can represent the result as a pure one-loop 
expression provided the corresponding formulae are written in terms 
of the {\em bare}, not physical, threshold scales. Thus, if 
the bare values of the thresholds are given (as would be the case if a 
grand unified theory emerges, say, as a limit of string theory) we 
could just forget about the 
second and all higher loops altogether.

Let 
me 
elucidate what I mean by this rather paradoxical statement.
As an example, consider the $SU(3)$ model described above, with 
one extra 
flavor in the fundamental representation (i.e. one chiral triplet 
superfield and 
one chiral antitriplet).  Since our task here is purely illustrative it is 
convenient to assume that the symmetry breaking $SU(3) 
\rightarrow U(1)^2$ 
takes place in two stages: first at a high scale $SU(3) \rightarrow 
U(1)\times 
SU(2) $  and then, at a somewhat lower scale, the remaining $SU(2)$ 
is broken 
to $U(1)$. In other words, instead of dealing with the general 
situation 
(which can be addressed, of course) we will assume that $|\varphi^8| 
\gg
|\varphi^3|$.  Among other simplifications this will allow us to 
neglect
the off-diagonal term 
$$
F_{\mu\nu}^{(3)}F_{\mu\nu}^{(8)}
$$
in the low-energy effective action. Indeed, the coefficient in front of 
this term
is
$$
\frac{3}{32\sqrt{3}\pi^2}\left(
\ln\frac{\varphi^8 -(\varphi^3/\sqrt{3})}{\varphi^8 
+(\varphi^3/\sqrt{3})}
\right) \, ;
$$
in the limit at hand and is suppressed as $|\varphi^3/\varphi^8|$.

Under the above pattern of  symmetry breaking the elephants
split into two groups -- four very heavy elephants, with mass 
squared
equal to $3|\varphi^8|^2/4$ emerging at the first stage and two not so 
heavy 
elephants with mass squared $|\varphi^3|^2$ emerging at the second 
stage.
There are four mass scales in the problem:
the ultraviolet cut off $M_0$, the elephant masses and the (bare)
matter mass $m_0$, with  the following hierarchy 
$$
m_0 \ll  |\varphi^3|\ll |\varphi^8| \ll M_0\, .
$$
As we descend  from $M_0$, we pass  the first elephant 
threshold where
the gauge group $SU(3)$ is broken down to $U(1)\times SU(2)$. 
The matter triplet becomes an $SU(2)$ doublet plus a singlet coupled 
only to 
the eighth photon. 
Then, below the second elephant threshold, only the Abelian 
subgroup 
survives;  we have two 
photons and three matter fields. The charges of these three matter 
fields
with respect to these two photons are
$$
\left( \frac{1}{2}\, , \,\,\,  -\frac{1}{2}\, , \,\,\,  0\right)\,\,\, 
\mbox{for the third photon,}
$$
and
\beq
\left( \frac{1}{2\sqrt{3}}\, , \,\,\,  \frac{1}{2\sqrt{3}}\, , \,\,\,  
-\frac{1}{\sqrt{3}}\right)\,\,\,
\mbox{for the eighth photon.} 
\label{charges}
\eeq
 The gauge couplings $\alpha_3$ and 
$\alpha_8$ 
continue to evolve due to the contribution of the matter fields. 

When we cross, in our descent, the first elephant threshold  we get 
two separate $\beta$ functions governing the evolution of 
$\alpha_8$ and the 
$SU(2)$ gauge coupling.  Then, at the second elephant threshold the 
$SU(2)$ 
gauge coupling becomes
$\alpha_3$. Below the matter threshold the evolution of the gauge 
coupling 
constants stops, and we arrive to what we call frozen couplings.

The final answer for the low-energy (frozen) coupling constants is 
known 
exactly, as was demonstrated above.
The all-order result for $\alpha_3$ was given in Eq. (\ref{alpha}).
For the sake of convenience we reproduce it here again, along with 
the answer 
for 
$\alpha_8$,
$$
\frac{1}{\alpha_3} 
=\frac{1}{\alpha_0}
-\frac{6}{2\pi} \ln\frac{M_0}{\left( \varphi^3_0\right)^{2/3}
\left(\sqrt{3} \varphi^8_0/2
\right)^{1/3}} + 
\frac{1}{2\pi}\ln\frac{M_0}{m_f^0}\, ,
$$
\vspace{0.3cm}
\beq
\frac{1}{\alpha_8} 
=\frac{1}{\alpha_0}
-\frac{6}{2\pi} \ln\frac{M_0}{\left(\sqrt{3} \varphi^8_0/2
\right)} + 
\frac{1}{2\pi}\ln\frac{M_0}{m_f^0}\, .
\label{2alphas}
\eeq
In the difference of the constants nearly everything drops out,
\beq
\frac{1}{\alpha_8}-\frac{1}{\alpha_3} = 
\frac{4}{2\pi}\ln 
\left(\frac{\sqrt{3}\varphi^8_0}{2\varphi^3_0}\right)\, .
\label{dif}
\eeq
I emphasize that it is the ratio of the fields' bare expectation values
 that enters here.
This result looks remarkably simple! One could superficially 
interpret it is follows.
At the first (heaviest) elephant threshold, 
$\sqrt{3}\varphi^8_0/2$, the couplings 
$1/\alpha$ are 
unified.
They diverge immediately, running linearly (in the log scale), 
according to the {\em 
one-loop} 
formula, all the way down. The slope of the linear running changes 
abruptly at 
the elephant thresholds. From 5 above the heaviest elephant 
threshold 
it changes to 3
for the $SU(2)$ coupling and $-1$ for the eighth photon.
Below the second elephant threshold, $\varphi^3_0$, 
where we deal with the $U(1) $ couplings, they run in parallel, in 
accordance 
with the charge assignment (\ref{charges}), until both couplings 
reach one 
and the same threshold at $m_0$, where they freeze at once (Fig. 1).
This  simple interpretation is operationally correct, but this is  
deceptive correctness, of course. 

\begin{figure}
  \input{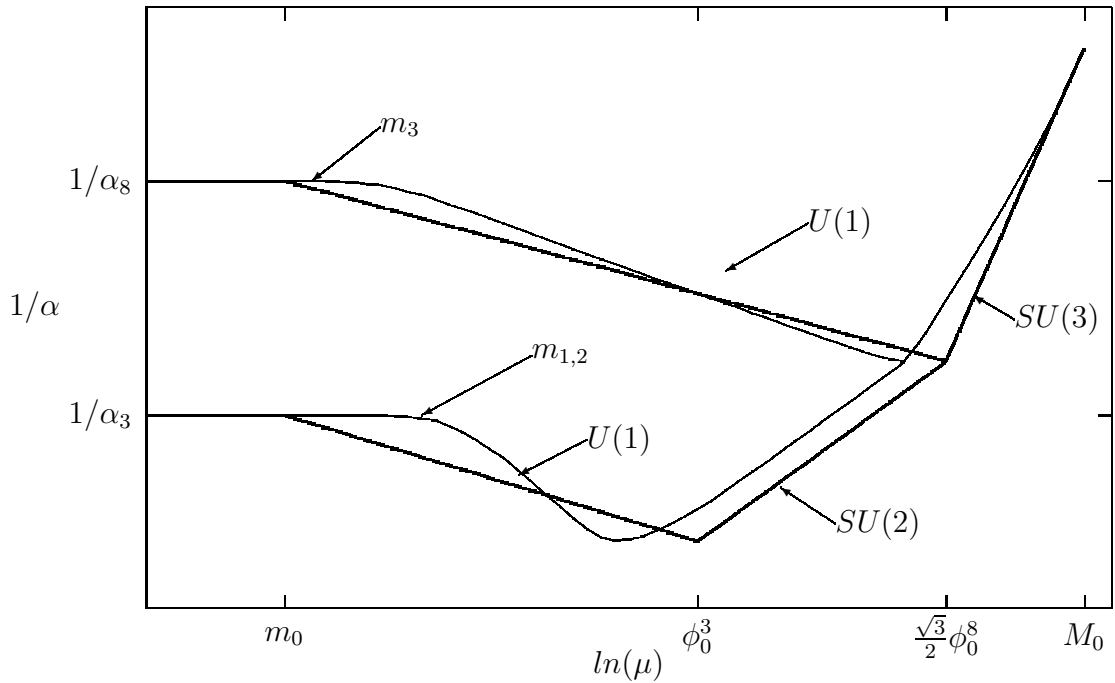}
  \caption{Evolution of the gauge couplings: from unification to 
freezing. The straight lines represent the one-loop evolution with the 
fake (bare) thresholds. The actual evolution is given by the smooth 
curves.}
  \label{fig1}
\end{figure} 

After briefly  reflecting one gets a second thought. First of all, it 
is 
absolutely 
certain that the $\beta$ functions are multiloop. Therefore, the 
running law is 
non-linear, and this non-linearity is different for $1/\alpha_3$ and
$1/\alpha_8$, in particular, in the interval between the two elephant 
thresholds since the $Z$ factors of the $SU(2) $ doublet  and singlet 
in this 
domain are definitely different. Second, for the same reason
the matter threshold can not be at one and the same place for the 
third and 
the eighth photons -- although we start from one and the same 
parameter
$m_0$ the physical masses of the three fermions we get after
$SU(3)$ is broken down to $U(1)^2$ are different. Therefore, the 
constants must
freeze
out at different scales. Finally,  $\sqrt{3}\varphi^8_0/2$ and 
$\varphi^3_0$ are not physical threshold scales as well; these are 
bare parameters, the corresponding physical thresholds differ by the 
$Z$ factors.

And yet, all these complications (the difference between the physical 
and bare thresholds and the higher orders in the $\beta$ functions) 
conspire together and kill each other
producing an effective picture of the one-loop evolution
with the slopes jumping at fictitious thresholds. The actual evolution 
can be obtained from the fictitious straight lines of Fig. 1 by slightly 
distorting the scale
in the horizontal direction. The distortion is logarithmic,
i.e. rather weak (provided all couplings enter in the weak regime
everywhere  below $M_0$) and is scale (and pattern)  dependent.
In the SU(3) and SU(2) parts we pull the thresholds to the left, while
in the U(1) parts we pull them to the right, by different amounts.
The distortions of the $\mu$ scale are determined by the $Z$ factors
which, although different from unity, are not strongly different. 
Under realistic conditions they may be 1.5 or 2 or so. 

Thus, we witness another little miracle. Although the derivation
presented above refers to a particular case, actually the message is 
general --
(i) use the bare values of the threshold parameters abstracted from 
the 
bare 
Lagrangian God-given at some high (say, Planck) scale; (ii) forget 
about the 
multiloop effects, just run and match the gauge couplings according 
to the 
most naive one-loop formula (the so-called $\vartheta$ function 
approximation). Then, at the very end, when the 
evolution is 
finished, you will get the correct full results for the low-energy 
gauge constants.

In particular, let us draw attention to the fact
that    the additional matter  produces no impact whatsoever in the 
difference of the couplings, Eq. (\ref{dif}). If we worked in terms of 
the physical thresholds, we would start from different physical 
masses for the three species of the additional matter.  The $Z$ factors 
will also be different. But the product of $m$'s times $Z$'s
will be the same. That's how the additional matter contribution will 
disappear in the difference of the gauge couplings, although it will be 
much harder to see this disappearance keeping in mind that the $Z$ 
factors are usually 
calculated approximately, not exactly. 

This feature is 
general. Say, in a realistic $SU(5)$ model all $SU(5)$ multiplets
provided  by a common mass parameter  are 
irrelevant for the 
calculation of the coupling differences. 

What will happen if we introduce physical threshold scales --
the ``GUT" scales $\sqrt{3}\varphi^8/2$, instead of the bare scales?
Then
$$\ln 
\left(\frac{\sqrt{3}\varphi^8_0}{2\varphi^3_0}\right)
\longrightarrow \ln 
\left(\frac{\sqrt{3}\varphi^8}{2\varphi^3}\right)
$$
plus logarithm of the $Z$ factor describing the evolution of the
$\phi$ field in the interval between the first and the second elephant 
threshold. If we transfer this logarithm from the right-hand side to 
the left-hand side of Eq. (\ref{dif}), on the left-hand side we will get 
a specific combination which is actually nothing else than the 
Wilsonean couplings \cite{SV2}. The issue of the thresholds will be 
discussed in more detail later on. 

\section{What if the evolution is not completed?}

The example discussed above is very elegant but, unfortunately,
is not very practical since usually the symmetry breaking pattern is 
such that not all gauge bosons acquire masses, so it is impossible to 
descend down,  below  the `` freezing point".  If the  freezing points 
existed, the scale where they  (the  freezing points) occur
is established as a result of a natural flow, and is different in 
different subgroups. This variation of scales neutralizes  the effect of 
the
second and higher-order terms in the $\beta$ function. 

In the case there is no freezing point, or we measure the running 
couplings above the freezing points, we usually measure them at a  
scale $\mu$ which is one and the same for all subgroups. Then 
the impact of the second and higher-order terms in the $\beta$ 
function is nonvanishing. To get a better idea of what is going on
we will consider another pedagogical example, the SU(3) model
we dwell upon above, but this time it will be assumed that the 
symmetry breaking pattern is $SU(3)\rightarrow SU(2)\times U(1)$.
Moreover, for the time being we will discard the additional matter, 
and add the
superpotential term to the $\Phi$ superfield
\beq
{\cal W} =\frac{1}{4g_0^2} \left(
m_0 \Phi^a \Phi^a + \lambda_0 d_{abc} \Phi^a\Phi^b\Phi^c
\right)\, ,
\label{supp}
\eeq
where $d_{abc} $ are the $d$ symbols of $SU(3)$. The superpotential 
(\ref{supp}) destroys the valley, and fixes the vacuum value of the
$\Phi$ field. One of the possible solutions
is
$$
\varphi_0^8 = \frac{2m_0}{\sqrt{3}\lambda_0} \, ;
$$
this solution is characterized by the property of vanishing 
$\varphi^3$
which ensures that the $SU(2)$ subgroup remains unbroken.  The 
bare mass squared of the  fourth, fifth, sixth and seventh gauge 
bosons, the elephants,  is 
$$
\left(  M^{(0)}_{4-7}\right)^2 = (3|\varphi_0^8|^2)/4
\, ;
$$
 they eat up the corresponding 
$\Phi$ superfields. 
The first three gauge bosons remain massless. There are two 
thresholds -- one associated with the elephants, another one 
with the fields $\Phi^{1,2,3}$. The latter  
 are massive because of 
the superpotential (\ref{supp}) and, in particular, 
the mass term of the $\Phi$ superfield. 
The bare mass of $\Phi^{1,2,3}$ is
$$
\left( M^{(0)}_{1-3} \right)^2= |m_0|^2 \, .
$$
 Let us assume that
the elephant threshold is heavier than that of  $\Phi^{1,2,3}$,
which is natural if the coupling constant $\lambda$ is weak.
Then below the lowest threshold associated with $\Phi^{1,2,3}$
we are left with  SU(2) supersymmetric gluodynamics,
with no matter fields, plus the eighth photon 
and no charged matter. So, the eighth gauge coupling is frozen;
the $SU(2)$ coupling, however, evolves till the very end --
this subgroup, being unbroken,  has no natural freezing point. 

Thus, far above the thresholds we start from  the unified gauge 
coupling
$\alpha_0$. Far below the  two thresholds we have
 the evolving coupling,
$\alpha_{\rm SU(2)}$, and the frozen coupling,  $\alpha_{\rm U(1)}$. 
Since 
the
$SU(2)$ coupling evolves we have to distinguish between the
Wilsonean constant -- it will be marked by  braces -- and that 
appearing in the $c$-number functional $\Gamma$ \cite{SV2}.
The Wilsonean constant $\{ 1/\alpha\}$ is renormalized only at one 
loop, while $1/\alpha$ is renormalized to all orders. The exact 
relation between $\{ 1/\alpha\}$ and $1/\alpha$ is explicitly known 
\cite{SV2}.

Let us choose the normalization point $\mu$ far below
the lowest threshold, $\mu \ll M_{1-3} $. We start from the following 
one-loop formula for the Wilsonean couplings (which is 
perturbatively exact)
$$
\left\{\frac{2\pi}{\alpha_{\rm SU(2)}(\mu )}\right\}
= \frac{2\pi}{\alpha_{0}} - 6 \ln\frac{M_0}{\mu} -
3\ln \frac{M_0}{(\sqrt{3}\varphi_0^8/2)} +
$$
\beq
2\ln\frac{M_0}{m_0} + \ln  \frac{M_0}{(\sqrt{3}\varphi_0^8/2)} 
\, ,
\label{asu2}
\eeq
and
\beq
\frac{2\pi}{\alpha_{\rm U(1)}} = 
\frac{2\pi}{\alpha_{0}} - 6 \ln\frac{M_0}{(\sqrt{3}\varphi_0^8/2)} \, 
.
\label{au1}
\eeq
The first line in Eq. (\ref{asu2}) reflects the contribution of the gauge 
bosons, and the fact that the fourth to seventh become elephants and 
freeze out. The second line in Eq. (\ref{asu2}) is  due to the $\Phi$ 
fields. The first term in the second line comes from $\Phi^{1,2,3}$, 
the second from
those $\Phi$'s ($\Phi^4$ to $\Phi^7$) which are eaten up by the 
elephants. We know already that Eqs. (\ref{asu2}) and
(\ref{au1}) are valid to all orders.

The expressions presented above are not so easy to grasp; they may 
even generate a couple of perplexing questions.
To get a transparent interpretation of this result  two further steps 
are in order.  First, we will pass from the Wilsonean coupling 
$\left\{1/{\alpha_{\rm SU(2)}(\mu )}\right\}$ to the regular one, 
$1/\alpha_{\rm SU(2)}(\mu )$. Second, 
it is convenient to consider the difference between 
$1/\alpha_{\rm SU(2)}(\mu )$ and $1/\alpha_{\rm U(1)}$, keeping 
in mind 
that the latter is frozen below the elephant threshold. Therefore,
$\alpha_{\rm U(1)}$ in the problem at hand is nothing else than
$\alpha_{\rm GUT}$. 

In pure SU(2) supersymmetric gluodynamics the first transition is 
realized as follows \cite{SV2}
\beq
\left\{\frac{2\pi}{\alpha_{\rm SU(2)}(\mu )}\right\}
= \frac{2\pi}{\alpha_{\rm SU(2)}(\mu )} + 2 
\ln \frac{\alpha_{\rm SU(2)}(\mu )}{\alpha_0}\, ,
\label{wils}
\eeq
where the factor 2 in front of the logarithm is actually $T(G)$ for
$G=SU(2)$.  In addition, we have to introduce the $Z$ factors 
describing the evolution of the $\Phi$ matter. Let $Z(M_0
\rightarrow M_{\rm GUT})$ express 
the evolution from $M_0$ down to the physical GUT scale $M_{4-7}
\equiv M_{\rm GUT}$, while
$Z(M_0
\rightarrow M_{1-3})$ expresses the evolution from $M_0$ down to 
the
physical $m$ threshold (for $\Phi^1$ to $\Phi^3$). Then, with our 
definitions, $m_0 =Z(M_0
\rightarrow M_{1-3})m$
and $\sqrt{3}\varphi_0^8/2 = M_{\rm GUT} [Z(M_0
\rightarrow M_{\rm GUT})]^{-
1/2}\alpha_0^{1/2} \alpha_{\rm GUT}^{-1/2}$.  Subtracting Eqs. 
(\ref{asu2}) and (\ref{au1}) from each other and using the physical 
thresholds we readily get
\beq
\frac{2\pi}{\alpha_{\rm SU(2)}(\mu )}=
\frac{2\pi}{\alpha_{\rm GUT}} -
6 \ln \frac{M_{\rm GUT}}{\mu \left( 
\frac{\alpha_{\rm GUT}}{\alpha_{\rm SU(2)}(\mu )}\right)^{1/3}}
+ 2\ln \frac{M_{\rm GUT}}{m Z(M_{\rm GUT}
\rightarrow M_{1-3})}
\,  ,
\label{corev}
\eeq
where the factor
$$
Z(M_{\rm GUT}
\rightarrow M_{1-3}) = Z(M_0
\rightarrow M_{1-3})/Z(M_0
\rightarrow M_{\rm GUT})
$$
expresses the evolution from the elephant (GUT) threshold to the 
$m$ threshold; thus, the product $m Z(M_{\rm GUT}
\rightarrow M_{1-3})$ is nothing else than  the 
mass parameter $M_{1-3}$ 
at the GUT scale (rather than the physical mass of $\Phi^{1-3}$).

Equation (\ref{corev}) has various appealing features. In particular,
the parameters $M_0$ and $\alpha_0$ referring to the  initial
scale, lying far above the GUT scale, drop out, as they  should.

Now, the last exercise aimed at  making our  toy model of grand 
unification a little bit more realistic. Let us add an extra matter 
superfield, say, in the fundamental representation of SU(3) (actually 
we will need two chiral superfields -- one triplet and another 
antitriplet), with no mass term. What happens then?

A qualitative impact is quite obvious -- the U(1) coupling constant
now runs below the GUT threshold, and never freezes.
Therefore, it is necessary to introduce now the Wilsonean coupling 
for U(1) as well. Quantitatively, an extra term appearing on the 
right-hand side of Eqs.  (\ref{asu2}) and (\ref{au1}) is
$$
\ln \frac{M_0}{\mu} \, ,
$$
one and the same in both expressions.  Simultaneously,
$2\pi / \alpha_{\rm U(1)}$ in Eq. (\ref{au1}) should be
substituted by $\{ 2\pi / \alpha_{\rm U(1)}\}$. Transition from the 
Wilsonean couplings to the regular ones is now  to be carried out in 
both subgroups, SU(2) and U(1). In the first case, instead of Eq. 
(\ref{wils}) we have
\beq
\left\{\frac{2\pi}{\alpha_{\rm SU(2)}(\mu )}\right\}
= \frac{2\pi}{\alpha_{\rm SU(2)}(\mu )} + 2 
\ln \frac{\alpha_{\rm SU(2)}(\mu )}{\alpha_0} +
\ln  Z(M_0\rightarrow\mu ; d.m.) \, ,
\label{wils1}
\eeq
where d.m. stands for the doublet matter (s.m. below stands for 
singlet matter, i.e. the matter field from the original triplet which 
does not belong to the SU(2) doublet). 

Furthermore,  for the U(1) subgroup
\beq
\left\{\frac{2\pi}{\alpha_{\rm U(1)}(\mu )}\right\}
= \frac{2\pi}{\alpha_{\rm U(1)}(\mu )} + 
\frac{1}{3}\ln  Z(M_0\rightarrow\mu ; d.m.) 
+\frac{2}{3}\ln Z(M_0\rightarrow\mu ; s.m.) \, .
\label{wils2}
\eeq
From  $M_0$ down to $M_{\rm GUT}$
the $Z$ factor is common, below the GUT scale it splits. The SU(2)
doublet interacts with both,  the SU(2) gauge bosons, and  the 
eighth photon (see Eq. (\ref{charges})).  The SU(2) singlet 
interacts 
only with the photon. 

Assembling all pieces together we observe  that after 
the additional matter is included, the following relation emerges 
$$
\frac{2\pi}{\alpha_{\rm SU(2)}(\mu )}-
\frac{2\pi}{\alpha_{\rm U(1)}(\mu )}  = 
$$
$$
-
6 \ln \frac{M_{\rm GUT}}{\mu \left(
\frac{\alpha_{\rm GUT}}{\alpha_{\rm SU(2)}(\mu )}\right)^{1/3}}
+ 2\ln \frac{M_{\rm GUT}}{m Z(M_{\rm GUT}
\rightarrow M_{1-3})}
+\ln\frac{M_{\rm GUT}}{\mu  Z(M_{\rm GUT}\rightarrow\mu ; d.m.) 
}- 
$$
\beq
\ln\frac{M_{\rm GUT}}{\mu  [Z(M_{\rm GUT}\rightarrow\mu ; 
d.m.)]^{1/3} 
[Z(M_{\rm GUT}\rightarrow\mu ; s.m.)]^{2/3}} \,  ,
\label{corevm}
\eeq
where each term in each line has a very clear physical interpretation.
The three terms in the second line are  due to the SU(2) gauge 
bosons,  
the heavy mass threshold
of the $\Phi$ fields; and the additional matter;
the third line is the evolution of the U(1) gauge coupling.

Our toy example is now generic enough in the sense that it includes 
all elements one may encounter in the realistic models. Therefore, at 
this 
stage we are able to formulate general conclusions. We start from the 
discussion of the mass thresholds.

\section{Mass thresholds}

The proper treatment of the heavy threshold effects in the evolution 
of the 
gauge couplings is crucial in establishing relations between the 
low-energy couplings following from the fact of unification at a 
higher scale. At the one-loop level there is no problem -- the so 
called $\vartheta$ function (or run-and-match) approximation is 
applicable. According to this approximation, as $\mu$ evolves from 
$M_{GUT}$ downwards, one just abruptly changes  the first 
coefficient 
of the $\beta$ function (i.e. the coefficient in front of the logarithm) 
at the physical value of the corresponding mass. Say, at the
$\Phi$ threshold, above $M_\Phi$ one includes in the coefficient
the $\Phi$ contribution while immediately below $M_\Phi$
one excludes the $\Phi$ fields altogether. 
At the one-loop level the $\vartheta$ function prescription is 
certainly 
correct.

The question arises beyond the leading log approximation where the 
above prescription is certainly incorrect. Following a tradition 
established in the pre-supersymmetric era the same $\vartheta$ 
function 
approximation is usually applied at two loops  \cite{thetaap}.  Even 
though numerically this may not  be such a  bad idea, 
it is interesting to ask what is the correct way of taking the
heavy mass thresholds in the gauge couplings into account.

An idea which immediately comes to one's mind is 
smoothing  the $\vartheta$ functions in some way \cite{smoothing}.
The smoothing procedure is rather ambiguous, it depends on 
conventions and, generally speaking, has no direct physical meaning.
This happens because the gauge couplings by themselves are not 
physical observables. The definition  we follow -- through 
the coefficient in front of $W^2$ in the effective action --
is directly related to observables (with the power accuracy) only 
provided we are far above the threshold or far below it. Fortunately, 
with respect to the heavy thresholds, we are always far below.
If we  calculate at low energies the effect due to the heavy fields
(which are integrated out since they can appear only in virtual loops)
in the {\it full}  theory it is all we need. And, as was shown above, 
that's exactly what we do.

Let us examine Eq. (\ref{corevm}) (the same conclusions follow
from the analysis of Sect. 3). The heavy threshold effect in this 
expression is
represented by the second term in the second line,
$$
 2\ln \frac{M_{\rm GUT}}{m Z(M_{\rm GUT}
\rightarrow M_{1-3})}\, .
$$
If the physical mass of the fields $\Phi^{1-3}$ coincides with
$M_{\rm GUT}$ (i.e. with the elephant masses), then the argument of 
the logarithm is unity -- there is no heavy threshold correction at all.
This is quite natural and uninteresting. What is more interesting is 
the 
case when the physical mass of the fields $\Phi^{1-3}$
is less than $M_{\rm GUT}$.  Then the threshold correction is 
non-vanishing. In the $\vartheta$ function approximation
the threshold correction will reduce to 
$$
 2\ln \frac{M_{\rm GUT}}{m}
$$
where $m$ is the physical mass of the fields $\Phi^{1-3}$
(the same as $M_{1-3}$). 

Now, the correct answer, taking into account  all orders, replaces the 
physical mass
$m$ by $m Z(M_{\rm GUT}
\rightarrow M_{1-3})$, or the corresponding {\it mass parameter at 
the
GUT scale}. If our research starts from  model-building and goes 
in 
the direction of phenomenology, the short distance (GUT) value of the 
mass parameter is primary, and no further  efforts are required. Case 
closed. If, however, we start from phenomenology, and operate
with the physical masses of the particles, then to properly include 
the
heavy threshold effects beyond one loop (i.e. beyond the $\vartheta$ 
function 
approximation) one has to calculate the corresponding $Z$ factors.
Unfortunately, the $Z$ factors are not calculable to all orders.
Note, however, that in the two-loop analysis of the gauge couplings 
the $Z$ factors appear only at one-loop  (i.e. in the leading log 
approximation). In other words, we reduce the level of complexity by 
one order.
In particular, if there are heavy threshold corrections in the $Z$ 
factors themselves ( they occur  more rarely than in the gauge 
couplings, if at all) it is legitimate  to account for them in 
the $\vartheta$ function approximation. This will provide us with 
the 
two-loop evolution of the gauge couplings properly accounting for 
the heavy thresholds. 

\section{General lessons and master formula}

In order to analyze the evolution of the gauge couplings  from
$M_{\rm GUT}$ down to some low scale,  first one starts from the 
standard one loop expressions written in the most primitive  
``run and match"   approximation, using the physical values of the 
mass parameters in the logarithms. By the mass parameters I mean 
the threshold masses, $M_{\rm GUT}$,  and the normalization point 
$\mu$. Then, in 
 the logarithms corresponding to the matter field contribution, 
substitute the mass parameter by the mass parameter multiplied
by the corresponding $Z$ factor. In the logarithms corresponding to 
the gauge field contribution, 
substitute the mass parameter by the mass parameter multiplied
by the corresponding $Z^{1/3}$ factor. For the gauge fields
the $Z$ factor is $1/\alpha$.  

The above procedure fully specifies the evolution process. It is 
perturbatively {\it exact to all orders},
and takes into account  the {\it heavy mass thresholds exactly}
provided that supersymmetry holds, i.e. we stay above the
scale of the supersymmetry breaking. 

(In practical calculations one usually limits oneself to  two-loop 
accuracy. Even for that limited purpose our master formula is 
extremely useful.  First, it spares one  the necessity of tabulating 
the two-loop coefficients of the $\beta$ function, a rather 
cumbersome task by itself. Only one loop coefficients enter the game, 
and all one loop coefficients are simple 
numbers having a geometric, and group-theoretic, meaning. Second, 
it provides with a consistent and exhaustive treatment of the mass 
thresholds, as was mentioned above. Third, numerical integration of 
the renormalization group 
equations becomes unnecessary. The final result is expressed in 
terms of the $Z$ factors. In the two-loop analysis the latter enter 
only at one loop level, or, more exactly, in the leading log 
approximation, which is equivalent to one loop.)

In the text book case of the SU(5) grand unification \cite{Moh}
the master formula is
$$
\frac{2\pi}{\alpha_3 (\mu)}
= \frac{2\pi}{\alpha_{\rm GUT}}
- 9 \ln \frac{M_{\rm GUT}}{\mu \left(
\frac{\alpha_{\rm GUT}}{\alpha_3(\mu )}\right)^{1/3}}+
\sum_{\rm gen}\left[
\ln \frac{M_{\rm GUT}}{\mu Z_{qL}}
+\frac{1}{2}\ln 
\frac{M_{\rm GUT}}{\mu Z_{UR}}+\frac{1}{2}\ln 
\frac{M_{\rm GUT}}{\mu Z_{DR}}\right]+
$$
\beq
+ 3 \ln \frac{M_{\rm GUT}}{M_\Phi}
+\ln \frac{M_{\rm GUT}}{M_{H_{u,d}}^{(3)}}\, ,
\label{al3}
\eeq

\vspace{0.3cm}

$$
\frac{2\pi}{\alpha_2 (\mu)}
= \frac{2\pi}{\alpha_{\rm GUT}}
- 6 \ln \frac{M_{\rm GUT}}{\mu \left(
\frac{\alpha_{\rm GUT}}{\alpha_2(\mu )}\right)^{1/3}}+
\sum_{\rm gen}\left[
\frac{3}{2}\ln \frac{M_{\rm GUT}}{\mu Z_{qL}}
+\frac{1}{2}\ln 
\frac{M_{\rm GUT}}{\mu Z_{\ell L}}\right]+
$$
\beq
2\ln \frac{M_{\rm GUT}}{M_\Phi} +
\ln 
\frac{M_{\rm GUT}}{M_{H_{u,d}}^{(2)}}\, ,
\label{al2}
\eeq

\vspace{0.3cm}

$$
\frac{2\pi}{\alpha_1 (\mu)}
= \frac{2\pi}{\alpha_{\rm GUT}} +
$$
$$
\frac{3}{10}\left[ \sum_{\rm gen}\left(
\ln \frac{M_{\rm GUT}}{\mu Z_{\ell L}}
+2\ln 
\frac{M_{\rm GUT}}{\mu Z_{e R}}+
\frac{1}{3}\ln 
\frac{M_{\rm GUT}}{\mu Z_{qL}}+
\frac{8}{3}\ln 
\frac{M_{\rm GUT}}{\mu Z_{UR}}+
\frac{2}{3}\ln 
\frac{M_{\rm GUT}}{\mu Z_{DR}}
\right)\right. +
$$
 \beq
\left.
2\ln 
\frac{M_{\rm GUT}}{M_{H_{u,d}}^{(2)}}
+\frac{4}{3}
\ln 
\frac{M_{\rm GUT}}{M_{H_{u,d}}^{(3)}}\right] \, .
\label{al1}
\eeq

Here I made various simplifying assumptions (which, in principle,
can be easily lifted).  First, it is  assumed, of course, that $\mu$
is larger than the superpartners' masses -- our master formula is 
valid only as long as supersymmetry holds. The masses of the two 
Higgs doublets of the model at hand, $M_{H_{u,d}}^{(2)}$  are also 
assumed to lie above 
the supersymmetry breaking scale, and above  $\mu$. 
Let us stress that $M_{H_{u,d}}^{(2)}$ in the expressions above is 
{\em not} the physical mass, 
but, rather the corresponding mass term at the GUT scale. 
At the  same 
time 
 the SU(2) breaking scale is taken to be below $\mu$ --
the $SU(3)\times SU(2)\times U(1)$ symmetry is manifest in Eqs.
(\ref{al3}) -- (\ref{al1}). 
 Moreover, the Yukawa interactions 
are neglected so that all three matter generations have identical $Z$ 
factors. This means that in the approximation accepted there are only 
five different matter $Z$ factors -- $Z_{qL}$ corresponding to the 
left-handed quark doublet, $Z_{UR}$ corresponding to the 
right-handed up-quark, $Z_{DR}$ corresponding to the 
right-handed down-quark, $Z_{\ell L}$ corresponding to the 
left-handed lepton doublet, and, finally, $Z_{e R}$ corresponding
to the right-handed electron (muon, $\tau$). $M_{H_{u,d}}^{(3)}$
stands for the mass term of the triplet superheavy Higgs particles 
(from $5$ and $\bar 5$),   as it is seen at the GUT scale.
 Finally, $M_\Phi$ is the
mass term of the chiral superfields from the $24$-plet at the GUT 
scale. 
The last line in each of the expressions above represents the 
GUT scale threshold terms, plus the doublet Higgs contribution in Eqs. 
(\ref{al2}), (\ref{al1}). 

All the simplifying assumptions listed above can be lifted in an 
obvious way. 

\section{$Z$ factors}

The master formula above expresses the gauge couplings, to all 
orders, in terms of the $Z$ factors. Unfortunately, the latter are not 
calculable to all orders. The $Z$ factors appear as the coefficients in 
front of the $D$ terms, and the theorems of the holomorphic 
dependence on the moduli and the (inverse) gauge couplings are not 
valid \cite{footn}. Without 
the power of holomorphicity one can not establish 
the all-order results for the $Z$ factors and has to resort to the 
old-fashioned order-by-order perturbative calculations. The 
anomalous dimensions of the matter fields were calculated up to
three loops \cite{3loops}. The  higher-order terms
in $\gamma$'s are definition-dependent, however, starting from the 
second loop. The work of
bringing the results in line with the definition where the NSVZ 
$\beta$ function  is valid is under way now \cite{KJ}. After this work 
is done 
one will be able
to calculate the gauge coupling evolution up to  fourth order. 

As was mentioned more than once, for the two-loop analysis
of the gauge couplings, it is sufficient to have the $Z$ factors
in the leading log approximation, which is unambiguous and 
scheme-independent. Actually, the one loop expressions can be 
obtained with 
{\it no calculations}, from general arguments alone.

The relevant one-loop diagram is presented on Fig. 2. Let us assume 
at first that we have only one chiral matter superfield, in the adjoint 
representation, with no self-interaction. Then the theory has $N=2$ 
supersymmetry, and the
$\beta$ function is well-known to be one-loop. From Eq. 
(\ref{NSVZbeta}) we then immediately read off that
$$
\gamma = -N\frac{\alpha}{\pi}\, .
$$
From the structure of the graph depicted on Fig. 2 it is clear that
the factor 
$N$ above is nothing else than $C_2({\rm adj})$
where 
$C_2$ is the  quadratic Casimir operator,
$$
\left( T^aT^a\right)_{{\rm rep}\,\,  R} = C_2(R){\bf 1}\, ,
$$
related to the Dynkin index in a standard way,
$$
C_2(R) = T(R)\frac{{\rm dim}\,\, ({\rm adj})}{{\rm dim}\,\, (R)}\, .
$$
At the one loop order the diagram of Fig. 2 is the only relevant 
graph. It is obvious then that 
in the general case of the matter field in the arbitrary representation 
of the gauge group, $C_2({\rm adj})$ is merely substituted by
$C_2(R)$, irrespective of how many matter fields we have, whether 
or not there are Yukawa interactions (self-interactions), and so on. 
We conclude that the one-loop gauge contribution to $Z$ is
\beq
Z = 1 - C_2 (R) \frac {\alpha_0}{\pi}\ln \frac{M_0}{\mu }\, .
\label{ZF}
\eeq
It is clear that the gauge-interaction   contribution is negative, 
tending to make $Z$'s less than unity.

\begin{figure}
  \epsfxsize=12cm
  \centerline{\epsfbox{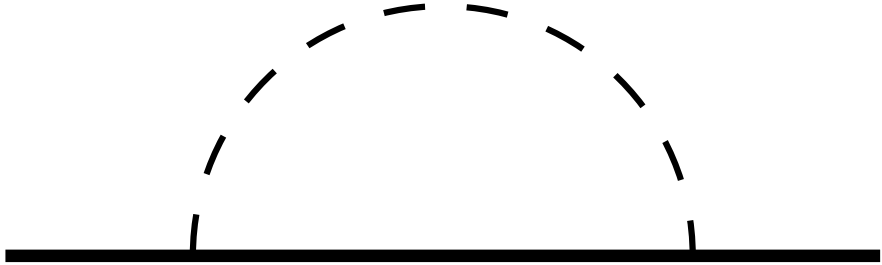}}
  \caption{ One-loop graph describing the matter field 
renormalization due to the exchange of the gauge field.}
  \label{fig2}
\end{figure}

The renormalization group improvement of expression
(\ref{ZF}) (summation of the leading logarithms) is trivial,
\beq
Z \rightarrow \left( \frac{\alpha_0}{\alpha} \right)^{2C_2/b_0}
\, ,
\label{Zimpr}
\eeq
where 
$$
b_0 = 3T(G) -\sum_i T(R_i)
$$
is the first coefficient in the $\beta$ function. 

Similar ``back-of-the-envelope" arguments yield the (super)Yukawa 
vertices' contribution to $Z$'s, which, on the contrary, tends to make
the $Z$ factors larger than unity. Indeed, let us start again from the
case of  adjoint matter, this time assuming that there are three
matter fields, $\Phi_1^a$, $\Phi_2^a$, and $\Phi_3^a$. If the Yukawa
coupling 
\beq
\Delta{\cal W} =\frac{1}{g_0^2} f^{abc}\Phi_1^a \Phi_2^b \Phi_3^c
\label{4extended}
\eeq
is added to the superpotential
the theory becomes $N=4$ supersymmetric, the gauge coupling does 
not run, and the anomalous dimensions of all $\Phi$ fields vanish.
This means that the renormalization of, say,  the $\Phi_1$ field
due to the gauge boson (Fig. 2) is canceled by a similar diagram
with the gauge boson line replaced by that of the $\Phi$ fields,
and the gauge vertices replaced by the (super)Yukawa vertices.
We immediately conclude then that
\beq
Z_{\rm Yukawa}  = 1 + N\frac {\alpha_0}{\pi}\ln \frac{M_0}{\mu }\, .
\label{ZYukawa}
\eeq
Getting the Yukawa contributions from Eq. (\ref{ZYukawa})
for other matter fields and different structure of the Yukawa vertices 
is a mere substitution of obvious color factors. 

It is convenient to collect here the leading-logarithm expressions for 
the
five $Z$ factors entering our master formula in the SU(5) case,
$$
Z_{qL} = Z_{\rm SU(3)}\times Z_{\rm SU(2)}\times  Z_1^{1/9} \, ,
$$
$$
Z_{UR} = Z_{\rm SU(3)}\times Z_1^{16/9} \, ,
$$
$$
Z_{DR} = Z_{\rm SU(3)}\times Z_1^{4/9} \, ,
$$
$$
Z_{\ell L} = Z_{\rm SU(2)}\times Z_1 \, ,
$$
and, finally,
\beq
Z_{eR} =  Z_1^{4} \, ,
\label{allz}
\eeq
where
\beq
Z_{\rm SU(3)} =\left[\frac{\alpha_{\rm GUT}}{\alpha_3 
(\mu)}\right]^{8/9}\, ,\,\,\,
Z_{\rm SU(2)} =\left[\frac{\alpha_{\rm GUT}}{\alpha_2 
(\mu)}\right]^{-3/2}\, ,\,\,\,
Z_1 =\left[\frac{\alpha_{\rm GUT}}{\alpha_1 (\mu)}\right]^{-1/22}
\, ,
\label{ququ}
\eeq
and the Yukawa interactions are 
 neglected, for simplicity.
The Yukawa interactions might be important  for the $t$ quark, but 
even in this case,
numerically, their effect is very modest. Moreover, it is 
assumed
 that the heavy thresholds coincide 
with the GUT scale, while the doublet Higgs particles are lighter
than $\mu$. It is quite clear how to amend this expression
if $\mu$ is smaller than the doublet Higgs masses. 

\section{Numerical analysis: bird's eye view}

Now, we will discuss the numerical situation, in gross features, 
leaving aside fine details. Our task is merely to get a general idea.
To this end I will neglect the low-energy thresholds due to 
supersymmetry breaking, assuming that all superpartners
have masses in the ballpark of 100 to 200 GeV.  Making some of 
them 
significantly heavier (say,  $\sim 1$ TeV) may change our 
conclusions.  This is a different topic, however, 
which I will not touch upon, referring the reader to the very rich 
literature devoted to the issue of  low-energy thresholds.
Two other possibilities -- new physics in the middle of the
great desert, and Planck physics corrections at the GUT scale --
will not be considered as well.  The first topic was  
covered in the talks of B. Brachmachari, E. Keith and some others,
while the second topic was extensively reviewed by Prof. P. Nath. 

To begin with, let us truncate the expressions
(\ref{al3}) -- (\ref{al1}) at one loop and disregard the heavy 
thresholds (i.e. assume that the masses of the Higgses from the 
24-plet and triplets from the quintets coincide with $M_{\rm GUT}$).
Then for the low-energy value of $\alpha_3$ we get
\beq
\frac{2\pi}{\alpha_3} =
\frac{12}{7} \, \,  \frac{2\pi}{\alpha_2} -
\frac{5}{7} \, \,  \frac{2\pi}{\alpha_1} \, ,
\label{number1}
\eeq
where by low energy I mean that the corresponding normalization 
point is somewhere in the ballpark of 100 to 200 GeV, see above. 
The value of $2\pi/\alpha_2$ is close to 186 and that of 
$2\pi/\alpha_1$ is close to 371. Where exactly the normalization 
point lies affects the numbers above at the level $\sim$ one unit;
a similar uncertainty is associated with the low-energy thresholds
(provided that the sparticle masses are $\sim$ 100 GeV).  The effects 
I am hunting for now constitute seven units -- this is the impact of 
the difference
between $\alpha_s (M_Z) =0.11$ and $\alpha_s (M_Z) = 0.125$.

Using Eq. (\ref{number1}) we conclude that $2\pi/\alpha_3
\approx 54$ (and $2\pi/\alpha_{\rm GUT} \approx 153)$. 
If the second loop shifted this number  up by three units, to 57, I 
would be very happy.
The above value then would   nicely match  what is expected 
from the low-energy QCD phenomenology, $\alpha_s (M_Z)
\approx 0.11$ \cite{Shif1}.  Unfortunately, the higher-order 
corrections work just in the opposite direction  shifting 
$2\pi/\alpha_3$  down by three units.  

By inspecting Eqs. (\ref{al3}) -- (\ref{al1}) and (\ref{allz})
we immediately observe that,  by far, the most significant numerical 
effect of higher loops comes from the factors
$$
\left( \frac{\alpha_{\rm GUT}}{\alpha_3(\mu )}\right)^{1/3}\,\,\,
\mbox{and} \,\,\, 
\left( \frac{\alpha_{\rm GUT}}{\alpha_2(\mu )}\right)^{1/3}
$$
in the logarithms in Eqs. (\ref{al3}), (\ref{al2}) and from 
$Z_{\rm SU(3)}$. The factors $Z_1$ and $Z_{\rm SU(2)}$
in the standard version of the SU(5) grand unification 
are too close to unity to be of importance in my
back-of-the-envelope analysis -- they are 
buried in the noise of other insignificant  corrections. Keeping only 
the important factors we get, instead of Eq. (\ref{number1}),
$$
\frac{2\pi}{\alpha_3} =
\frac{12}{7} \,\,  \frac{2\pi}{\alpha_2} -
\frac{5}{7}  \,\, \frac{2\pi}{\alpha_1} +
$$
\beq
3\ln \left( \frac{\alpha_{\rm GUT}}{\alpha_3(\mu )}\right)
-\frac{24}{7}\ln \left( \frac{\alpha_{\rm GUT}}{\alpha_2(\mu 
)}\right)
-\frac{9}{14} \ln Z_{\rm SU(3)} \, . 
\label{number2}
\eeq
In the second line we can use the values of the constants
in the one-loop approximation; moreover,  $Z_{\rm SU(3)}$
is given by Eq. (\ref{ququ}). The most significant contribution, $-3$,  
comes from the  first term in the second line. The second  and the 
third terms are close to 0.6 and nearly compensate each other.
As a result, the second loop shifts the prediction for
$2\pi/\alpha_3$ 
away from the right value, from 54 to 51 (i.e. $\alpha_s (M_Z) 
\approx 0.123$). 

Since we are so  remarkably close to the desired number
it is quite natural to think that the residual discrepancy is
due to some corrections that are unaccounted for so far.
The idea which immediately comes to one's mind
is including the Yukawa interactions in order to change the values
of the $Z$ factors. We -- Ian Kogan and myself -- played with this 
idea for some time. To achieve success in this way one has to make 
an  appropriate $Z$ factor of order 100  or larger ($\ln Z \sim 5$). It 
is quite obvious that in the minimal model even the strongest
Yukawa interaction, that of the $t$ quark, is far too weak
to ensure such a large value of $Z$. Only if some Yukawa constant 
approaches the Landau pole can we expect to get $Z$ that large.

Another way out may be associated with the heavy particles' 
thresholds. Generally speaking, the masses of the Higgs particles 
from the 24-plet and the triplets from the quintets need not exactly 
coincide with $M_{\rm GUT}$, see the second lines in Eqs.  
(\ref{al3}) -- (\ref{al1}). Again, it is clear that to shift the prediction 
for $2\pi/\alpha$ by several units  the logarithm of the mass ratio 
has to be several units itself. In other words, the superheavy Higgs 
particles have to be hundred  times lighter than $M_{\rm 
GUT}$. In this case there is the  menace of a fast proton decay due to 
dimension five operators \cite{D5}.
In more details the scenario was investigated in Refs. \cite{M,B,L}. It 
was shown that in the SU(5) grand unification this
solution does not work, unfortunately.  The proton decays too fast 
\cite{M,B}.
In the SO(10) grand unification, however, the superheavy Higgs 
thresholds do bring the value of $\alpha_s (M_Z)$ down to 0.11
without making the proton lifetime unacceptably short \cite{L}.
Moreover, the prediction for the proton lifetime returns
to the range accessible experimentally, which makes the whole story 
exciting.

\section{Conclusions}

$\bullet$  \hspace{0.5cm} If the primary ``high energy"    gauge 
group 
$G$  is spontaneously broken, $G\rightarrow G_1\times 
G_2\times ...$, then the low-energy couplings $1/\alpha_1$,
$1/\alpha_2$, ... are given by a master formula 
{\it to all orders}, including mass thresholds.

\vspace{0.3cm}

\noindent $\bullet$  \hspace{0.5cm}  If supersymmetric mass 
parameters are 
known at the GUT scale
(rather than the physical mass parameters) then full inclusion of the 
threshold effects trivializes.  

\vspace{0.3cm}

\noindent $\bullet$  \hspace{0.5cm} If the evolution is complete (i.e. 
all
$G_i$'s  are  $U(1)$'s,   and the low-energy couplings are frozen, then 
the 
all-order result for $1/\alpha_i$ looks as if it was one-loop, with {\it 
faked} (bare) thresholds.

\vspace{0.3cm}

\noindent $\bullet$  \hspace{0.5cm}
The center of gravity of the multiloop analyses is shifted
from the gauge $\beta$ functions to the $Z$ factors of the matter 
fields. The complexity (number of loops) is reduced by one level.
In the minimal SU(5) grand unification the dominant two-loop 
correction comes from the one-loop $Z$ factor of the gluon (gluino) 
field.
It is of the right magnitude, but, unfortunately, its sign is 
``unfavorable",  so 
that the value of $\alpha_s (M_Z)$ becomes unacceptably high. 

\section{Acknowledgments}

 I would like to thank
I. Kogan, L. Roszkowski, and A. Vainshtein for stimulating 
discussions. 
Assistance of B. Chibisov  with plots is gratefully 
acknowledged. This work was 
supported 
in part by  
DOE under the grant number
 DE-FG02-94ER40823.


\begin{thebibliography} {99}
\bibitem{exact}
If all mater fields we have in the model belong to the
fundamental representation of the gauge group it is possible to prove 
that the NSVZ $\beta$ function is exact also including
nonperturbative effects. The proof can be abstracted from Ref. 
\cite{SV3}.
 In this work the gaugino
condensate
$\langle\lambda\lambda\rangle$ was exactly calculated. Since
$\langle\lambda\lambda\rangle$ is in the same superfield as the 
anomaly
of the trace of the stress tensor
$\langle\lambda\lambda\rangle$ has zero anomalous dimension; if it 
is
known exactly the full $\beta$ function immediately follows.
\bibitem{footn}
Perhaps, some elements can be maintained, in a 
certain limited sense. The hope is not yet lost  completely.
\end{thebibliography}
\end{document}